# Influence of graphene-substrate interactions on configurations of organic molecules on graphene: pentacene/epitaxial graphene/SiC

W. Jung,[1] D.-H. Oh,[1] I. Song,[1] H.-C. Shin,[1] S. J. Ahn,[1] Y. Moon,[1] C.-Y. Park,[1, a)] and J. R. Ahn[1,2, b)]

[1)]*Department of Physics, Sungkyunkwan University, Suwon 440-746, Republic of Korea*

[2)]*SAINT, Sungkyunkwan University, Suwon 440-746, Republic of Korea*

(Dated: 16 May 2014)

Pentacene has been used widely in organic devices, and the interface structure between pentacene and a substrate is known to significantly influence device performances. Here we demonstrate that molecular ordering of pentacene on graphene depends on the interaction between graphene and its underlying SiC substrate. The adsorption of pentacene molecules on zero-layer and single-layer graphene, which were grown on a Si-faced 6H-SiC(0001) wafer, was studied using scanning tunneling microscopy (STM). Pentacene molecules form a quasi-amorphous layer on zero-layer graphene which interacts strongly with the underlying SiC substrate. In contrast, they form a uniformly ordered layer on the single-layer graphene having a weak graphene-SiC interaction. Furthermore, we could change the configuration of pentacene molecules on the single-layer graphene by using STM tips. The results suggest that the molecular ordering of pentacene on graphene and the pentacene/graphene interface structure can be controlled by a graphene-substrate interaction.

[a)]Electronic mail: cypark@skku.edu

[b)]Electronic mail: jrahn@skku.edu

Graphene is an extraordinary material exhibiting unique physical properties that impart it with excellent prospects for a variety of applications.[1–11] Among them, in the field of electronic devices, graphene has replaced electrode and/or channel parts to enhance device performances.[8,10–13] For example, graphene was used as a channel material in a radio frequency field emission transistor (FET)[9] and as an electrode material in a solar cell[12] and a light emission diode (LED).[13] Furthermore, one of the technical advances in electronic devices has been the development of organic devices that are lighter, more flexible, and less expensive than inorganic devices.[14,15] Therefore, it has only been logical to utilize graphene in organic devices such as organic FETs, solar cells, and LEDs.[16–18] One of the key factors that determine the performance of organic devices is the interface structure between the organic molecules and the electrode materials.[16–18] In the development of graphene-based organic devices, understanding and controlling the configuration of organic molecules on graphene is one of the essential focal points of research.[16–18] Among the various organic materials, pentacene has been regarded as one of the most promising candidates for an organic FET because of its high carrier mobility, chemical stability, and compatibility with low-temperature Si fabrication processes.[17,18] In earlier studies, gold was used as an electrode material in a pentacene FET, but showed a significant charge-injection barrier because of an unfavorable interface dipole layer formation.[17,18] In contrast to the gold electrode, when graphene was employed as the electrode of a pentacene FET, a superior interface contact between graphene and pentacene reduced the contact resistance and the charge-injection barrier height.[17,18]

Therefore, it is necessary to understand and manipulate the interface structure between organic molecules and graphene, to enhance further the performance of graphene-based organic devices. On metal substrates, it was reported that the orientation of a pentacene molecule could depend on the electronic structure of the substrate, when Bi(001) and Au(111) surfaces were used as the substrate.[19] For pentacene on exfoliated single-layer and bilayer graphene on a silicon dioxide ($SiO_2$) film, it was reported that the activation energy for molecular aggregation depended on a water layer at the graphene/$SiO_2$ interface.[20] A systematic investigation is thus required to study molecular ordering of pentacene on graphene with different atomic and/or electronic structures. An extreme case is to use metallic and insulating graphene. Both insulating and metallic graphene can be epitaxially grown on 6H-SiC(0001).[3,5,7,21,22] Zero-layer graphene on 6H-SiC(0001) has the same atomic structure as single-layer graphene but is insulating because most of the carbon atoms of the zero-layer graphene are covalently bonded to the Si atoms of the underlying SiC substrate.[21,22] In contrast, single-layer graphene located on the zero-layer graphene is metallic, showing typical Dirac electron behavior.[21,22]

Here we report that the molecular ordering of pentacene on the zero-layer graphene is quite different from that on the single-layer graphene, as observed by scanning tunneling microscopy (STM). On the insulating zero-layer graphene, pentacene molecules at low pentacene coverage were adsorbed with three preferential orientations at room temperature (RT)[23], but did not show a long-range order, resulting in a quasi-amorphous phase at high pentacene coverage. In contrast, pentacene molecules on the metallic single-layer graphene at low pentacene coverage were mobile at RT and formed a uniformly ordered layer at high pentacene coverage. The results suggest that when the interaction between graphene and its underlying substrate is changed, the atomic and/or electronic structures of graphene get modified thereby impacting the molecular ordering of pentacene on graphene and hence the resulting pentacene/graphene interface. Furthermore, because the interaction between pentacene and the single-layer graphene is weak, as can be concluded from the uniform ordering of the pentacene on single-layer graphene, we could change this ordering and the orientation of the pentacene layer using an STM tip.

Epitaxial graphene was grown on a Si-faced 6H-SiC(0001) wafer in an ultra-high vacuum (UHV) chamber with a base pressure of $5 \times 10^{-11}$ Torr. The SiC wafer was hydrogen-etched in a separate chamber before the growth of graphene. The hydrogen-etched SiC wafer was then transferred to the UHV chamber and heated overnight to 600 ℃ to degas. The wafer was next exposed to Si flux while being maintained at 850 ℃, to remove native Si oxides. The zero-layer graphene was grown after heating the wafer to 1150 ℃, whereas the single-layer graphene was grown after heating the wafer to 1200 ℃. Pentacene in a graphite crucible was outgassed overnight in the UHV chamber. Pentacene flux was controlled by adjusting the temperature of the crucible. All pentacene films on graphene in this report were grown at RT, where the pressure of the chamber was maintained below $2 \times 10^{-10}$ Torr during the deposition of pentacene. All STM images were acquired at RT using an Omicron VT-STM in constant current mode, where electrochemically-etched tungsten tips were employed.

Figure 1(a) and 1(b) show STM images of the superstructures of the zero-layer and single-layer graphene, respectively. When a Si-faced 6H-SiC(0001) wafer was heated to 850 ℃ under Si flux, its surface reconstructed to a Si-terminated (3 × 3) phase.[24] The Si-terminated surface further reconstructed sequentially to and $(6\sqrt{3} \times 6\sqrt{3})R30°$ when heating to higher temperatures.[24] The $(\sqrt{3} \times \sqrt{3})R30°$ surface is another Si-terminated structure consisting of Si adatoms while the $(6\sqrt{3} \times 6\sqrt{3})R30°$ surface is C-terminated.[24] The $(6\sqrt{3} \times 6\sqrt{3})R30°$ superstructure is nothing but the zero-layer graphene having the same atomic structure as single-layer graphene, as shown in Figure 1(c).[22,24] Most of the carbon atoms of the zero-layer graphene are covalently bonded to the Si atoms of the underlying SiC substrate, resulting in an insulating grapheme, not showing Dirac electron behavior.[22,24] The SiC wafer was heated to a higher temperature of 1200℃ to grow single-layer graphene, as displayed in Figure 1(b).[24] The single-layer graphene has a (6 × 6) superstructure and is located on the zero-layer graphene [Figure 1(d)].[21,22,24] The single-layer graphene shows typical linear energy dispersions of Dirac electrons and is n-type because electrons are transferred to it from the SiC substrate.[21,22,24]

Pentacene molecules were deposited on the zero-layer and single-layer graphene at RT. Figure 2(a) shows an STM image of pentacene-covered zero-layer graphene at low pentacene coverage. A "kidney bean"-like feature in Figure 2(a) represents a single pentacene molecule lying flat, with the inset displaying the molecular structure of pentacene overlapped with the "kidney bean" like feature. This feature of pentacene in an STM image is consistent with STM images of pentacene on other substrates.[23,25,26] The image shows that the single pentacene molecule has three preferential orientations, as indicated by yellow arrows, which is in accordance with STM experiments reported previously.[23] The existence of the three preferential orientations suggests there is a strong interaction between the zero-layer graphene and the single pentacene molecule. In the zero-layer graphene [Figure 1(c)], there are two kinds of carbon atoms: one has an unsaturated $\pi$ bond and the other has a saturated $\pi$ bond.[22,24] The carbon atoms with unsaturated $\pi$ bonds may contribute to the adsorption of pentacene on the zero-layer graphene because of their chemical reactivity, resulting in three preferential orientations of pentacene following the structural symmetry of the zero-layer graphene, where the yellow hexagons indicate the $(6\sqrt{3} \times 6\sqrt{3})R30°$ superstructure. When pentacene coverage is increased, the molecules begin to pair, as indicated by dotted circles in Figure 2(b). The intermolecular distance of a pentacene molecule pair was approximately 9.77 Å The intermolecular distance is more than that found in a pentacene bulk phase,[27] which suggests that the intermolecular interaction of pentacene molecules on the zero-layer graphene is weaker than the interaction between pentacene and the zero-layer graphene. At higher pentacene coverage, the three preferential orientations of pentacene molecules were maintained and the adsorbed molecules were not mobile at RT as a result of which the pentacene molecules did not show any long-range order, as shown in Figure 2(c).

In contrast to the zero-layer graphene, the growth mechanism of pentacene molecules on single-layer graphene was much different, as displayed in Figure 3 and 4. On the single-layer graphene, at low pentacene coverage, single pentacene molecules were not observed in the STM images at RT. The nonexistence of single pentacene molecules in STM images does not imply that they were not adsorbed on the single-layer graphene because, at high coverage, pentacene molecules were observed in the STM images. The nonexistence of single pentacene molecules in STM images thus suggests that single pentacene molecules are very mobile at RT on the single-layer graphene, as opposed to those on the zero-layer graphene. As mentioned before, however, single pentacene molecules were observed on the zero-layer graphene at RT. The disparity in behavior of the pentacene molecules at low and high coverages between the zero-layer and single layer graphenes lead us to conclude that the interaction between pentacene and single-layer graphene is much weaker than that between pentacene and zero-layer graphene. When pentacene molecules fully covered the single-layer graphene, resulting in the first pentacene layer, they could be observed in the STM images [Figure 3(a)]. The domains of the zero-layer and single-layer graphene in Figure 3(a) are indicated by domains I and II, respectively. The domains can be clearly determined because the configurations of pentacene molecules are much different from each other: pentacene molecules are disordered in domain-I but show long-range order in domain-II. The domains I and II can be thus assigned to zero-layer and single-layer graphene, respectively. The configuration of pentacene molecules on the single-layer graphene (domain-II) resembles an array of linear molecular chains, as seen in Figure 3(d), where the molecular structure of pentacene is shown to overlap the enlarged STM image. In this configuration, the pentacene molecules are orientated along the chain direction and the inter-chain distance is approximately 11.9 Å The preferential direction of the edges of the epitaxial graphene domains on 6H-SiC(0001) was reported to be the armchair direction,[28] while the graphene domains on metal substrates prefer zigzag edges.[29] The red arrows in Figure 3(a) indicate the armchair directions. The orientation of the pentacene chains on the metallic single-layer graphene can thus be assigned a zigzag direction, as indicated by the yellow arrow in Figure 3(a).

Because of the weak interaction between pentacene and single-layer graphene, we could remove pentacene molecules at RT using an STM tip, as shown in Figure 3(b) and 3(c). In the course of repeated scanning with a bias voltage ($V_s$) of -2.7 V, pentacene molecules were gradually removed. Interestingly, the pentacene molecular chains experienced selective removal: every other pentacene molecular chain was removed, as shown in Figure 3(b). As a result, the inter-chain distance of the pentacene molecular chains increased from 11.9 to 19.2 Å The selective removal of the molecules suggests there is another ordered phase of pentacene molecules at a lower pentacene coverage. The orientations of the pentacene molecules were tilted at an angle in the molecular chains with the wider inter-chain distance, as shown in the enlarged STM image [Figure 3(e)]. The inter-chain distance of 19.2 Å is similar to the size of the (6 × 6) unit cell of the single-layer graphene. The stability of the pentacene molecular chains can therefore be related to an interaction between pentacene and the superstructure of the single-layer graphene. In one case, we could even change the orientation of the molecular chains, as shown in Figure 3(c). The orientation of the lower pentacene molecular chains were rotated by 60°, as indicated by the yellow arrow, while the upper chains maintained the same orientation.

After the first pentacene layer fully covered the single-layer graphene, the second pentacene layer began to grow, as can be seen in Figure 4. In Figure 4(a), domain-I indicates pentacene-covered zero-layer graphene and domain-II and domain-III indicate the first and the second pentacene layers on single-layer graphene, respectively. As described above, the domain of the pentacene-covered zero-layer graphene can be clearly distinguished by the disordered configuration of pentacene molecules (domain-I). Furthermore, the first pentacene layer on single-layer graphene can be also ascertained by the linear chain configuration of pentacene molecules (domain-II). However, this feature of the linear molecular chains was not as clearly visible as that observed before the growth of the second pentacene layer. This may have been caused by the presence of mobile pentacene molecules on the first pentacene layer, during the growth of the second pentacene layer. The second pentacene layer (domain-III) began to grow from the edges of singe-layer graphene domains [Figure 4(a)] and fully covered the first pentacene layer [Figure 4(b)]. Interestingly, the second pentacene layer grew continuously through step edges, as indicated by the dotted rectangle in Figure 4(b). The orientation of the pentacene molecules in the second layer is similar to that of the first pentacene layer on graphite.[30] The pentacene molecule is tilted at an angle with respect to the chain direction [Figure 4(c) and 4(d)]. The unit cell of the second pentacene layer is indicated in Figure 4(e). The inter-chain distance, indicated by *a*, is approximately 17.7 Å and the intermolecular distance along the chain direction, indicated by *b*, is approximately 7.0 Å, where the angle γ is approximately 73°.

In conclusion, the effect of the atomic and/or electronic structures of graphene on the growth of pentacene was studied using STM. The atomic and electronic structures of graphene were changed by the interaction between graphene and its underlying substrate. The zero-layer of epitaxial graphene on 6H-SiC(0001) was selected to behave as insulating graphene, where chemically reactive carbon atoms with unsaturated π-bonds coexist with carbon atoms having saturated π-bonds. The single-layer graphene on 6H-SiC(0001) was chosen to behave as metallic graphene showing typical Dirac electron behavior. Pentacene-graphene interaction was found to be strong when adsorbed on the zero-layer graphene but weak on the single-layer graphene. On the zero-layer graphene, there are preferential adsorption sites at which pentacene molecules are immobile at RT. The immobile pentacene molecules have local preferential orientations but do not show a long-range order. On the single-layer graphene, however, the pentacene molecules are very mobile at RT, resulting in uniformly ordered pentacene layers at high pentacene coverage. Therefore, we suggest that the configuration of pentacene molecules on graphene can be controlled by a graphene-substrate interaction. The results of our study will pave the way for the development of a designed interface structure between organic molecules and graphene in graphene-based organic devices.

This study was supported by a National Research Foundation of Korea (NRF) grant (No. 2012R1A1A2041241).

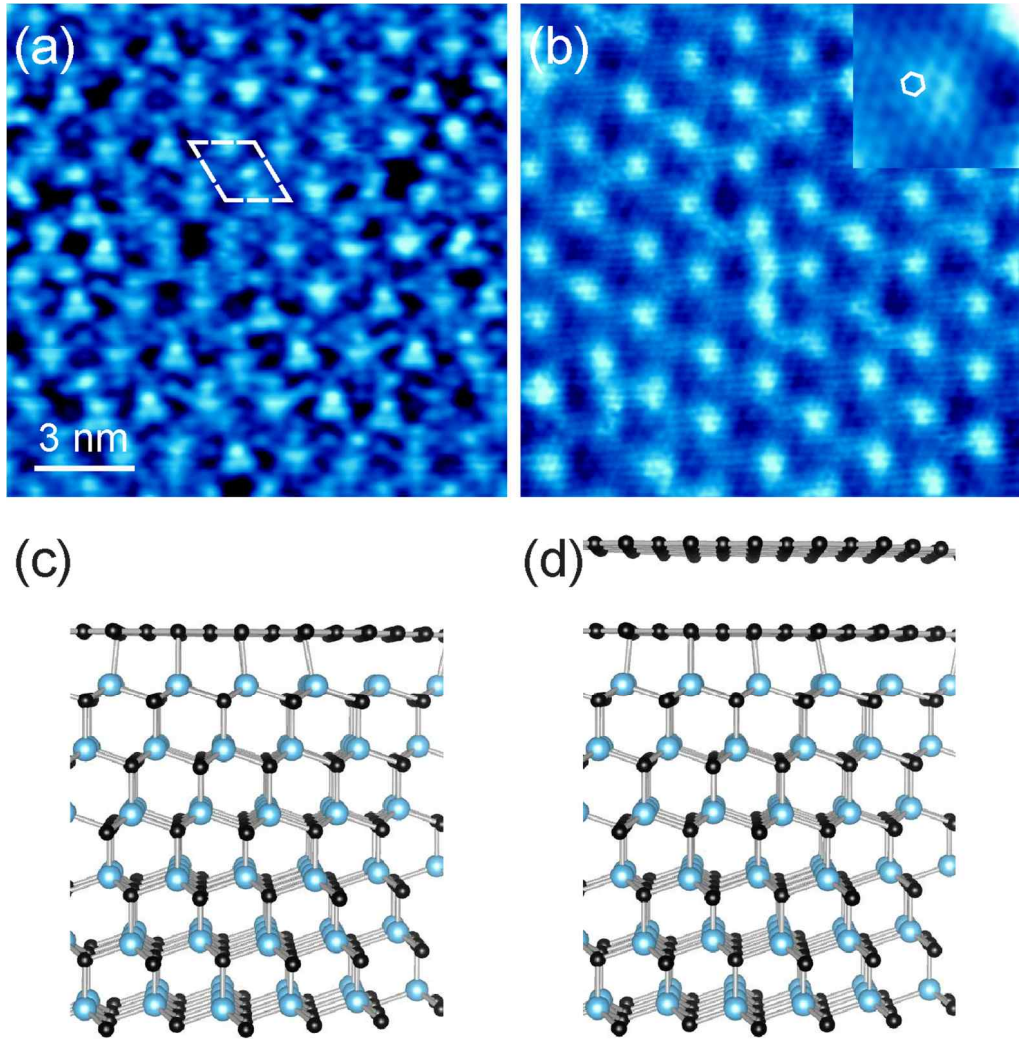

FIG. 1. (Color online) STM images of (a) zero-layer graphene ($V_s = -1.5$ V, tunneling current ($I_t$) = 100 pA) and (b) single-layer graphene ($V_s = -0.1$ V, $I_t = 500$ pA). Dashed lines in (a) indicate the superstructure of the zero-layer graphene. The inset in (b) is an enlarged STM image, where the solid hexagon indicates the atomic hexagonal ring of graphene. Atomic structure models of (c) zero-layer and (d) single-layer graphene on 6H-SiC(0001), where blue and black spheres indicate Si and C atoms, respectively.

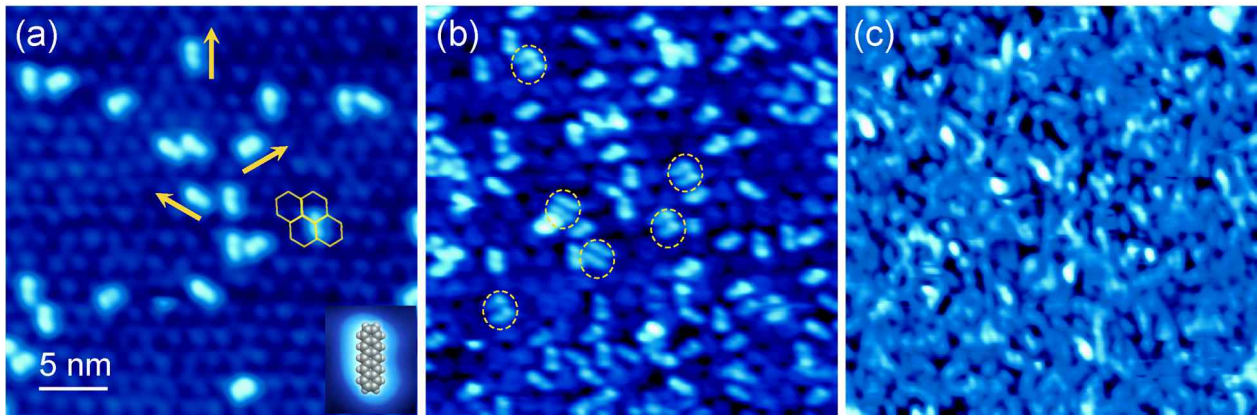

FIG. 2. (Color online) STM images of pentacene covered zero-layer graphene with different pentacene coverages. (a) An STM image ($V_s = -3.0$ V, $I_t = 100$ pA) with low pentacene coverage, where yellow arrows indicate the molecular orientations of pentacene, the yellow hexagons indicate the superstructure of the zero-layer graphene, and the molecular structure of pentacene is overlapped with its STM image. (b) An STM image ($V_s = -2.1$ V, $I_t = 50$ pA) with intermediate pentacene coverage, where the dotted yellow rings indicate pentacene molecule pairs. (c) An STM image ($V_s = -2.1$ V, $I_t = 50$ pA) with high pentacene coverage.

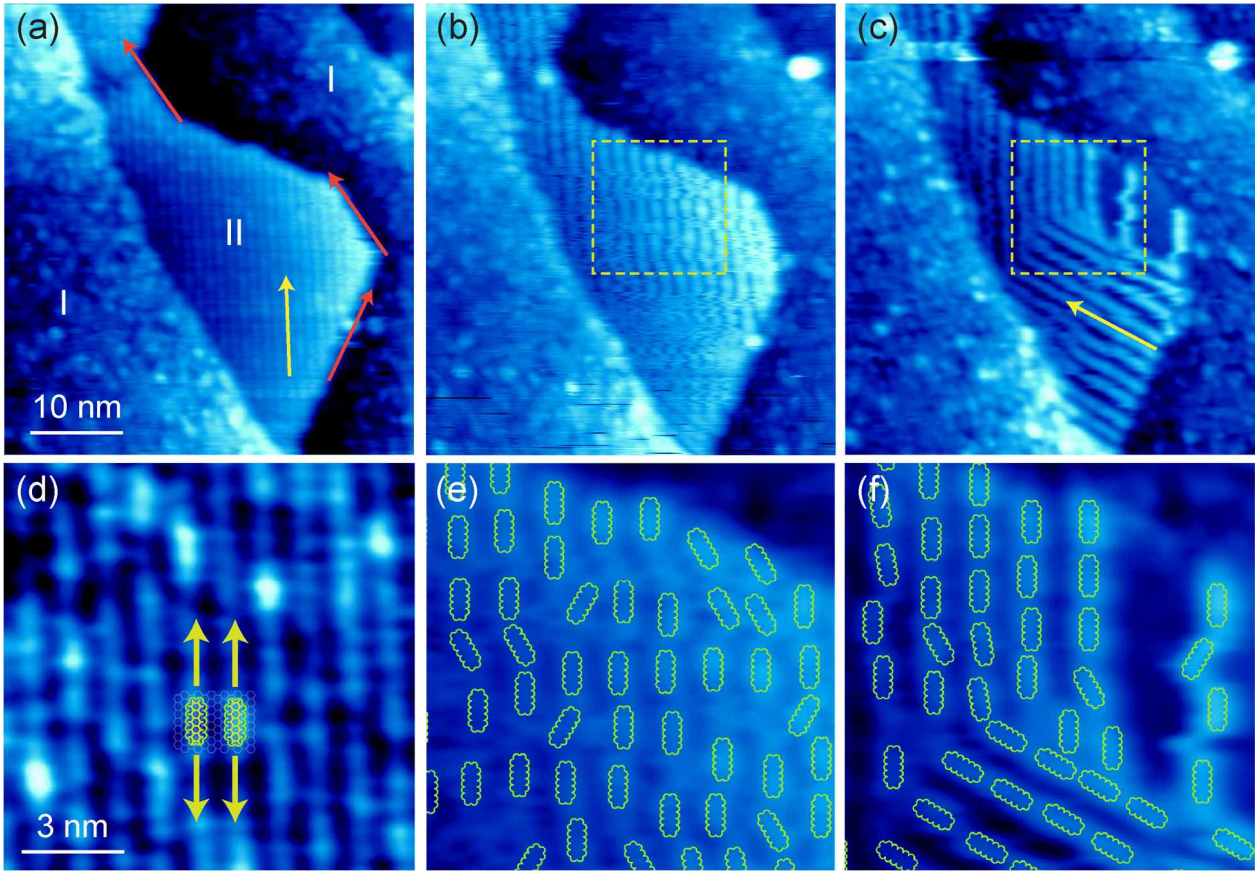

FIG. 3. (Color online) (a)-(c) STM images of the first pentacene layer on single-layer graphene. (a) An STM image of the saturated first pentacene layer ($V_s = -2.7$ V, $I_t = 30$ pA), where I and II indicate the domains of the zero-layer and single-layer graphene, respectively and the red and yellow arrows indicate the directions of the edge and the pentacene molecule chain, respectively. (b)-(c) STM images ($V_s = -3.0$ V, $I_t = 30$ pA) acquired after removing pentacene molecules by using an STM tip. (d)-(f) Enlarged STM images of (a), (b), and (c), respectively, where pentacene molecules are indicated by yellow shapes.

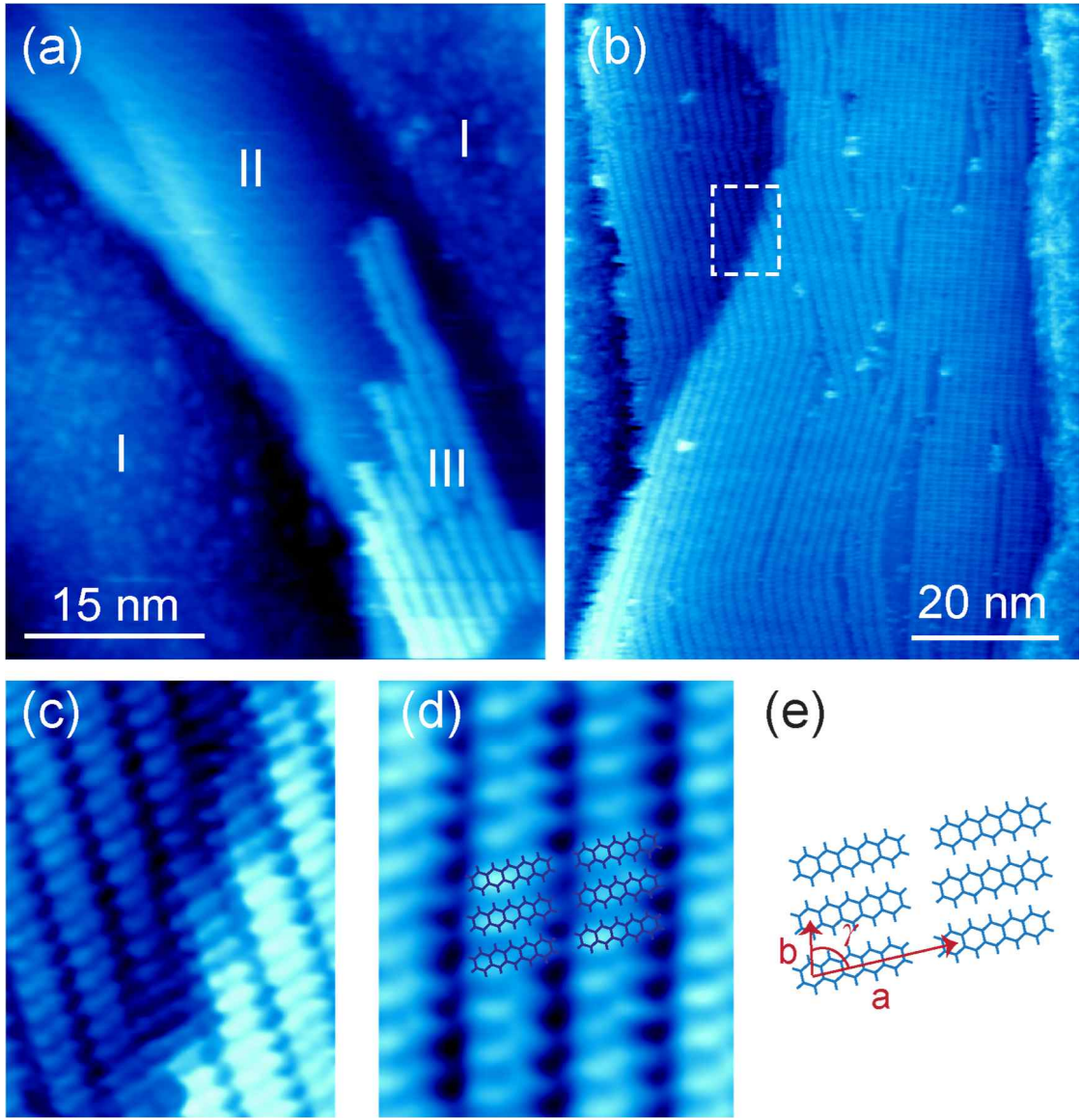

FIG. 4. (Color online) (a) An STM image ($V_s = -3.0$ V, $I_t = 30$ pA) of the second pentacene layer on single-layer graphene, where I, II, and III indicate pentacene-covered zero-layer graphene and the first and the second pentacene layers on single-layer graphene, respectively. (b) An STM image ($V_s = +1.6$ V, $I_t = 30$ pA) of the fully covered second pentacene layer on single-layer graphene. (c)-(d) Enlarged STM images of (b). (e) The unit cell of the second pentacene layer.